\documentclass[prl,aps,twocolumn,preprintnumbers, showpacs, nofootinbib,superscriptaddress,notitlepage]{revtex4-1}

\usepackage{mathrsfs}
\usepackage{amsfonts}
\usepackage{amsmath}
\usepackage{slashed}
\usepackage{array}
\usepackage{verbatim}
\usepackage{epsfig}
\usepackage{graphicx}
\usepackage{color}
\usepackage[dvipsnames]{xcolor}
\usepackage{multirow}
\usepackage{booktabs}
\usepackage{mathrsfs}
\usepackage[normalem]{ulem}

\newcommand{\beq}{\begin{eqnarray}}
\newcommand{\eeq}{\end{eqnarray}}

\begin{document}

\title{Shedding light on the intrinsic transversal momentum distributions of pion and kaon}

\author{Jian Chai}
\affiliation{School of Physics and Electronics, Hunan University, 410082 Changsha, China}
\affiliation{Institute of Theoretical Physics, School of Sciences,
Henan University of Technology, Zhengzhou, Henan 450001, China}
\author{Shan Cheng}\email{Corresponding author: scheng@hnu.edu.cn}
\affiliation{School of Physics and Electronics, Hunan University, 410082 Changsha, China}
\affiliation{School for Theoretical Physics, Hunan University, 410082, Changsha, China}
\affiliation{Hunan Provincial Key Laboratory of High-Energy Scale Physics and Applications, 410082 Changsha, China.}

\date{\today}

\begin{abstract}

We propose to introduce the intrinsic transversal momentum distribution functions (iTMDs), 
in conjunction with the light-cone distribution amplitudes (LCDAs), 
to elucidate the probability amplitude of encountering a meson state wherein the partons swiftly traverse along the longitudinal axis 
while gently oscillating in the transversal plane. 
The primary motivation stems from the oversight of soft transverse dynamics within the $k_T$ factorization formalism of an exclusive QCD process, 
which confines perturbative QCD (pQCD) predictions to scenarios involving large momentum transfers.
We meticulously investigate the $\pi$ and $K$ electromagnetic form factors using the iTMDs-improved pQCD calculation at next-to-leading order. 
By analyzing data in the timelike physical regions, we obtain the transversal-size parameters 
$\beta_\pi^2 = 0.51 \pm 0.04$ GeV$^{-2}$ and $\beta_K^2 = 0.30 \pm 0.05$ GeV$^2$. 
We then extract the chiral mass of pion to be $m_0^\pi(1 \, {\rm GeV}) = 1.84 \pm 0.07$ GeV 
and explain the precise measurements of kaon form factor in the perturbative timelike region. 
As a remarkable byproduct, we found that the incorporation of iTMDs improves the pQCD predictions for electromagnetic form factors, extending the applicable range to a few GeV$^2$. 
This improvement allows for direct comparison with existing measurements and lattice QCD evaluations.

\end{abstract}

\maketitle

\textbf{\textit{Introduction.}}--The form factor is a momentum-dependent function that encapsulates the characteristics of a specific interaction 
within an appropriate matrix element. 
Its measurement serves a crucial purpose in either validating or refining theories, 
such as perturbative QCD (pQCD) and quark confinement \cite{Mueller:1981sg,PQCD}. 
Pion and kaon electromagnetic form factors, defined by the simplest hadronic matrix elements, 
occupy a central role in the exploration of hadron structure and the breaking of chiral symmetry \cite{Efremov:1979qk,Lepage:1980fj}. 

The pQCD prediction of electromagnetic form factors is written in a factorization formalism \cite{Cheng:2019ruz}
\beq &&{\cal F}(Q^2) = \sum_{t_1,t_2} \int du_i d b_i \, \varphi^{(t_1)}(u_1, \mu_{r_1}) \otimes \varphi^{(t_2)}(u_2, \mu_{r_2}) \nonumber  \\
&& \hspace{1cm} \otimes {\cal H}^{(t_1 t_2)}(u_i, b_i,\mu_f) \times e^{-S(u_i, b_i, \mu_f)}.  \label{eq:ff-fact-kt} \eeq
In the formalism, contributions from different twists light-cone distribution amplitudes (LCDAs) of initial and final pions, 
denoted by $\varphi^{t_1}$ and $\varphi^{t_2}$ respectively, are summed up to give the total result. 
LCDAs at a certain twist $\varphi$ is a function of the longitudinal momentum fraction $u$ carried by the antiquark inside the meson 
and the renormalization scale $\mu_r$. 
${\cal H}^{(t_1 t_2)}$ are the perturbatively calculable hard kernels ordered by twists too, 
it indicates the probability of momenta redistributions at quark-gluon level due to the electromagnetic interaction happed within a small distance. 
Besides the momentum fraction $u_i$ and the hard factorization scale $\mu_f$, 
it depends also on the coordinate $b$ conjugated to the transversal momentum $k_T$, 
which is introduced in the energetic propagator to regulate the end-point singularity. 
Generally speaking, the transversal momentum varies within three scales, 
namely the QCD scale $\lambda$, the hard-collinear scale $\sqrt{\Lambda Q}$ and the hard scale $Q$. 
The loop integral of $k_T$ results in several large logarithms, especially in the integral regions $k_T \sim \Lambda$, 
they are subsequently resummed up to the well-known sudakov factor $e^{-S}$.
This exponent turns out to suppress the soft contribution, meanwhile highlight the hard scattering mechanism. 

The nonperturbative LCDAs are universal objects defined by matrix elements \cite{Braun:1989iv,Braun:2003rp,Ball:2006wn}, 
\beq \langle 0 \vert {\bar u}(z) \Gamma[z, 0] d(0) \vert \pi^-(p) \rangle \propto \int du e^{i u p \cdot z} \varphi(u, \mu_r), \label{eq:LCDAs} \eeq
where the nonlocal quark current with a suitable Dirac matrix $\Gamma$ is sandwiched between $\pi$ and the QCD vacuum, 
the path-ordered gauge factor (abbreviated as square bracket) is along the the light cone $z^2 = 0$. 
LCDAs describes the probability amplitudes to find the $\pi$ in a state 
with minimal number of constituents and have small transversal separation of order $1/\mu_r$. 
It is usually obtained by utilizing the conformal symmetry in massless QCD. 
Keep in mind that the conformal symmetry of a quantum theory implies a vanished $\beta$-function, 
hence it is powerful used in a QCD process with large momentum transfers and large energies. 
In practice, the renormalization scale is commonly chosen to be equal to the factorization scale $\mu_r = \mu_f$. 
So roughly speaking, the LCDAs embodied in the factorization formalism are the wave functions at zero transverse separations, 
this means that the soft transversal dynamics is actually missed in the factorization formalism shown in Eq. (\ref{eq:ff-fact-kt}). 

In the previous pQCD study of pion form factor \cite{Chai:2023htt}, 
the chiral mass $m_0^\pi(1 \, {\rm GeV}) = 1.37 \pm 0.30$ GeV is obtained 
by fitting the next-to-leading-order (NLO) prediction to the data-driven spacelike form factor. 
The result is much smaller than that obtained from chiral perturbative theory (ChPT) $1.89$ GeV \cite{Leutwyler:1996qg}
and from the ${\rm \overline{MS}}$ current quark mass $m_0^\pi = m_\pi^2/(m_u+m_d) \sim 1.74$ GeV \cite{ParticleDataGroup:2024cfk}. 
This deviation necessitates a reevaluation of the pQCD factorization formalism, 
the next-to-next-leading-order (NNLO) corrections and the high twist contributions are the first considerations. 
We convince ourself that the NNLO would not the resolve of the $m_0^\pi$ problem 
since the NLO calculations \cite{Li:2010nn,Cheng:2014gba} have showed a good perturbative convergence (about $10 \%$ increases). 
In addition, besides the chiral enhancement from the two-particle twist three LCDAs, 
higher twist contributions are expected to be tiny due to the heavy power suppression \cite{Cheng:2020vwr}. 
In these senses, the $m_0^\pi$ problem reveals a possible large effect from the transversal momentum distribution inside pions
when they are moving outside the scope of hard scattering. 

\begin{figure}[h] \begin{center} 
\includegraphics[width=0.40\textwidth]{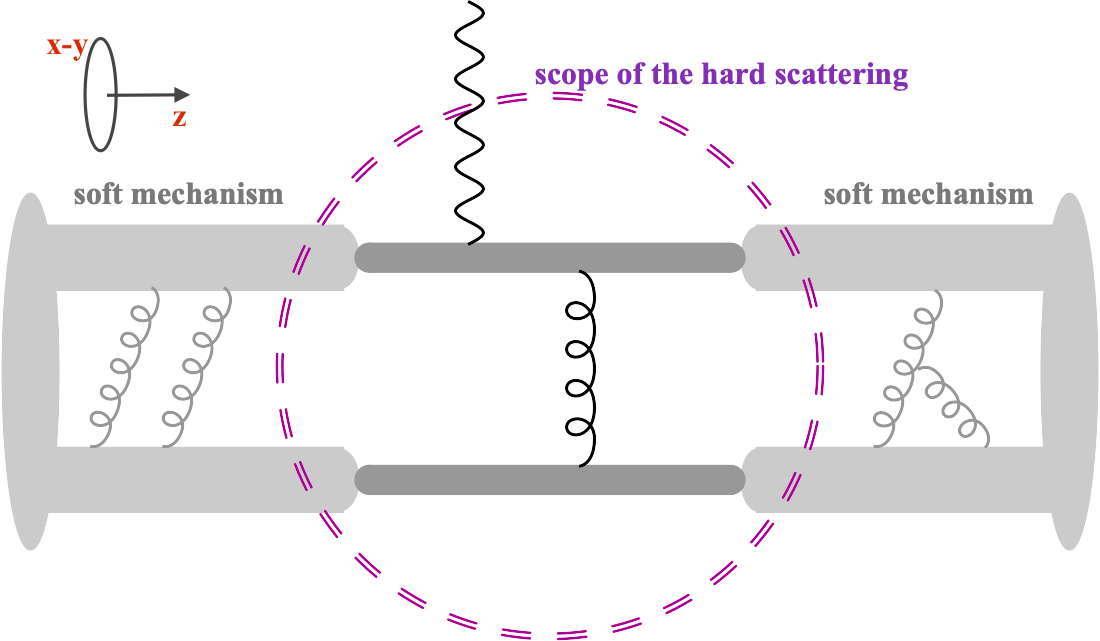} 
\end{center} \vspace{-6mm}
\caption{The sketch map of pion electromagnetic form factor.}
\label{fig:kTfactorization}
\end{figure}

In figure \ref{fig:kTfactorization} we show the sketch map of electromagnetic interaction of pion. 
The strength of electromagnetic interactions is divided into two distinct regions by a double-dashed-curves. 
Within the circle lies the realm of hard scattering (central region of electromagnetic potential field), 
described by the sudakov-multiplied hard kernel ${\cal H} e^{-S}$,
which picks up the hard radiations of partons on the transversal plane, as shown by the thin gray stick. 
Outside the scope of hard scattering, energetic pions are moving fast along the $z$ direction accompanied by soft bremsstrahlung radiations. 
The bremsstrahlung radiations along the light-cone is absorbed into the effects of high twist LCDAs, 
in terms of a new coordinate $x^2$ that approximates but does not exactly coincide with the light-cone coordinate. 
In the factorization formalism Eq. (\ref{eq:ff-fact-kt}), the hard transversal dynamics of LCDAs has been defined 
by selecting the renormalization scale of the nonperturbative parameters at the hard factorization scale
which aligns with the marginal region of the central potential field. 
In the exterior region, situated far from the central potential field, the soft radiations in the transversal plane 
(depicted by the thick light-gray stick) 
are notably absent from the definition of LCDAs and have often been overlooked in previous pQCD studies. 
This is the main motivation of this work. 

In this letter, we introduce the intrinsic transversal momentum distribution (iTMD) 
to describe the soft transversal degree of freedom associated to LCDAs.  
We ingeniously study the iTMDs shapes of pion and kaon via the electromagnetic form factors. 
The central goal is to mark up the missed soft transversal dynamics clarified above, 
hence to construct a comprehensive factorization for an exclusive QCD process. 

\textbf{\textit{Intrinsic transversal momentum dependent functions.}}--The soft pion wave function of the valence quark state 
is generally written by a product of LCDA and iTMDs  
\begin{widetext} \vspace{-3mm}
\beq &~&\langle 0 \vert {\bar u}(x) \Gamma[x^-, x_\perp; 0, 0_\perp] d(0) \vert \pi^-(p) \rangle 
\propto \int du d k^2_\perp e^{i u p^+ x^- - i k_\perp \cdot x_\perp} \psi(u, k_T), \quad 
\psi(u, k_T) = \frac{f_{\cal P}}{2\sqrt{6}} \varphi(u,\mu_r) \Sigma(u, k_T). \label{eq:pi-soft-DA} \eeq
\vspace{-5mm} \end{widetext}
They obey the normalization conditions \cite{LBHM}
\beq \int_0^1 du \varphi(u,\mu_r) = 1, \; \int \frac{d^2k_\perp}{16\pi^3} \Sigma(u,k_T) = 1. \label{eq:lt-twf-norm} \eeq
Since iTMDs encapsulate the soft transverse dynamics, deriving their expression from first principles is infeasible. 
Instead, we adopt a simple gaussian function to parameterize them while preserving rotational invariance \cite{Jakob:1993iw,Kroll:2010bf} 
\beq \Sigma(u, k_T) = 16 \pi^2 \frac{\beta^2}{u (1-u)} e^{ - \frac{\beta^2 k_T^2}{u (1-u)} }. \label{eq:TMD-lt-kt} \eeq 
It can be figurative considered by a transversal harmonic oscillator with the transversal-size parameter $\beta^2$. 
The transversal radius of the valence quark state should be no larger than the mean electric charge radius 
of the corresponding mesons, so the average value of the conjugate transversal momentum satisfies 
\beq \left[ \langle k_T^2 \rangle \right]^{\frac{1}{2}} \equiv 
\left[ \frac{\int du d^2 k_T k_T^2 \vert \psi(u, k_T) \vert^2}{\int du d^2 k_T \vert \psi(u, k_T) \vert^2} \right]^{\frac{1}{2}} 
\gtrsim \frac{0.2}{\langle r^2 \rangle^{\frac{1}{2}} }. \label{eq:iTMD-average} \eeq
Performing fourier transformation, we obtain the soft transversal function in the conjugated coordinate space 
\beq \hat{\Sigma}(u, b_T) = 4 \pi e^{- \frac{b_T^2 u (1 - u)}{4\beta^2} }.  \label{eq:TMD-lt-bt} \eeq 

\begin{figure*}[t] \begin{center}\vspace{-2mm}
\includegraphics[width=0.9\textwidth]{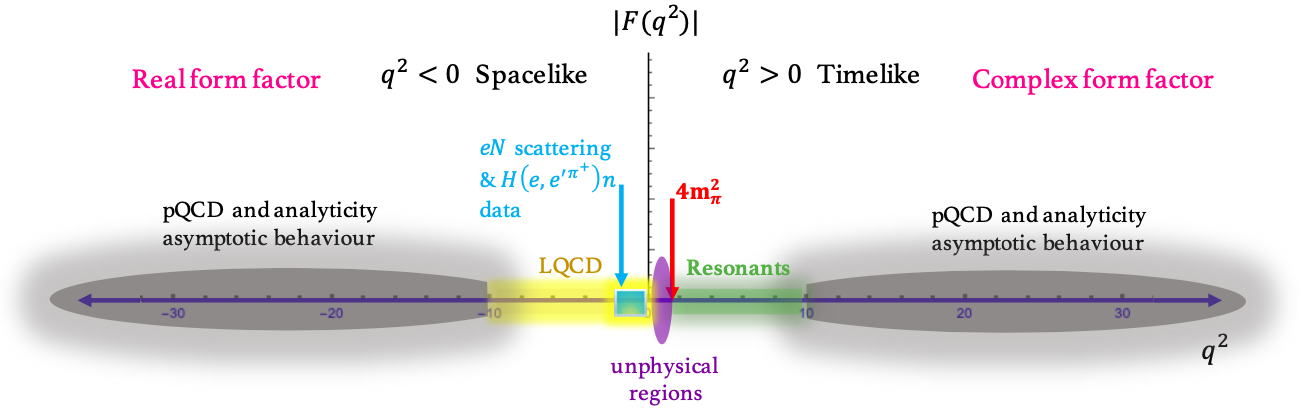}
\end{center}\vspace{-6mm}
\caption{Kinematical clarification of pion electromagnetic form factor.}
\label{fig:pi-ff-regions}
\end{figure*}

In the pQCD calculation, twist three LCDAs of valence quark state give the dominate contributions 
to pion and kaon electromagnetic form factors due to the chiral enhancement effect. 
Generally speaking, there are three sources for the high (power) twist LCDAs, say 
the "bad" components with "wrong" spin projection in the wave functions, 
the transversal motion of valence quark state in the leading twist components, 
and the higher Fock states with an additional gluon or quark-antiquark pair. 
The first two are defined by the genuine two-particle twist three LCDAs and their scale evolutions, 
meanwhile, the third one gives the quark mass correction terms in the two-particle twist three LCDAs. 
Note that the second source is partially related to the third one by the equation of motion. 
We hightlight that the transversal motion in the high twist LCDAs is definitely independent of the iTMDs introduced in this work, 
since they transfer different dynamics in the factorization formalism as stated at the end of last section. 

For the two-particle twist three LCDAs, we propose to introduce the iTMDs in the way 
\beq&&\psi^{p,\sigma}(u, \mu) = \int \frac{d^2 k_T}{16\pi^3}  \varphi_{2p}^{p,\sigma}(u, \mu) \Sigma(u, k_T) \nonumber \\
&&\hspace{0.2cm} + \int \frac{d^2 k_{1T} d^2 k_{2T}}{64\pi^5} \rho_+ \varphi_{3p}^{p,\sigma}(u, \mu) 
\int {\cal D} \alpha_i \Sigma^\prime(\alpha_i, k_{iT}). \label{eq:iTMD} \eeq
Here the superscript $p$ ($\sigma$) indicates the twist three LCDAs induced by the pseudoscalar (tensor) current, 
the subscripts $2p$ and $3p$ denote the contributions from valence quark state and three-particle state, respectively. 
The three-particle contribution is proportional to the quark mass, so a dimensionless quantity $\rho_+ = (m_{q_1} + m_{q_2})/m_0^{\cal P}$ 
is defined with the chiral mass $m_0^{\cal P}$. 
The integral measure over the longitudinal momentum fractions $\alpha_{i=1,2,3}$ 
reads as $\int {\cal D} \alpha_i = \int_0^u d \alpha_1 \int_0^{\bar u} d \alpha_2 \frac{1}{1-\alpha_1-\alpha_2}$. 
The normalization $\int_0^1 du \psi^{p,\sigma}(u, \mu) = 1$ is guaranteed by 
\beq &&\int \frac{d^2 k_{1T} d^2 k_{2T}}{64\pi^5} \int {\cal D} \alpha_i \Sigma^\prime(\alpha_i, k_{iT}) = 1, \nonumber\\
&&\int_0^1 du \, \varphi_{2p}^{p,\sigma}(u, \mu) = 1, \quad \int_0^1 du \, \varphi_{3p}^{p,\sigma}(u, \mu) = 0. \label{eq:norm-Sigma-prime}\eeq
Three-particle iTMDs is chosen again in a Gaussian function, in the conjugate coordinate space it reads as 
\beq \hat{\Sigma}^\prime(\alpha_i, b_{1}, b_{2}) = 4 \pi 
e^{- \frac{2 \alpha_3 (b_{1}^2 + b_{2}^2 ) + (\alpha_1 + \alpha_2) (b_{1} - b_{2})^2 }{16\beta'^2} }. \label{eq:TMD-t3-bt} \eeq
In order to derive Eq. (\ref{eq:TMD-t3-bt}), we use the crossing symmetry between $\alpha_1$ and $\alpha_2$. 
We see from Eq. (\ref{eq:iTMD}) that there are two iTMDs in the twist three soft wave function, 
corresponding to the two-particle and three-particle contributions. 
Two transversal-size parameters are hence involved, in which $\beta'^2$ is expected to be smaller than $\beta^2$ 
since the color charged soft gluon would shrink up the transversal extent of hadrons. 

\textbf{\textit{Electromagnetic form factors and pQCD prediction.}}--Electromagnetic form factor of pion has been investigated for several decades. 
It is a touchstone of QCD-based approaches, such as the Light-cone sum rules (LCSRs) \cite{Braun:1999uj,Bijnens:2002mg}, 
the Dyson-Schwinger equation (DSE) \cite{Chang:2013nia,Roberts:2021nhw}, and the pQCD \cite{Li:2010nn,Chai:2023htt} and et al., 
not only in their initial stage of establishment, but also in the well-developed period. 
An impressive study recently comes from the lattice QCD (LQCD), 
they improved the first-principle evaluation from the small momentum transfers $- 1 \, {\rm GeV}^2 \leq q^2 \leq 0$ \cite{Wang:2020nbf} 
to the large values $- 10 \, {\rm GeV}^2 \leq q^2 \leq 0$ \cite{Ding:2024lfj}. 
Meanwhile, the two-loop computation of leading-twist contribution in the hard-collinear factorization 
shows an enormous correction to the short-distance coefficient function \cite{Ji:2024iak,Chen:2023byr}. 
Among these approaches, the pQCD has an unique advantage to predict the form factors in both spacelike and timelike regions, 
what's more, it provides a systematical analysis of the contributions from different twist LCDAs.
From the experimental side, the precise data of pion electromagnetic form factor 
are mainly obtained in the physical regions by electron-position collider experiments \cite{BaBar:2012bdw,Belle:2008xpe,BESIII:2015equ}. 
In contrast, the data of spacelike form factor \cite{NA7:1986vav,JeffersonLabFpi-2:2006ysh,JeffersonLab:2008jve}, 
is available only in the small momentum transfers $[-2.50, -0.25]$GeV$^2$. 
In figure \ref{fig:pi-ff-regions}, we plot the research statue of pion electromagnetic form factor in different kinematical regions. 

People usually take a dispersion relation to connect the QCD predictions of spacelike from factor to the precise measurements in timelike region. 
In the standard representation, the spacelike form factor is written in an integral of the imaginary part of the timelike one, 
which is parameterized in a resonant model. 
Recently, we proposed the modular dispersion relation \cite{Cheng:2020vwr,Chai:2023htt}
\beq &~& \mathcal{F}_\pi (q^2<0) \nonumber\\
&=& \exp \left[ \frac{q^2 \sqrt{s_0 - q^2}}{2 \pi} \int\limits_{s_0}^\infty d s 
\frac{ \ln |\mathcal{F}_\pi (s)|^2}{s\,\sqrt{s - s_0}  \, (s -q^2)} \right].  \label{eq:Fpi_DR2} \eeq
In the modular representation, the spacelike form factor is written in an integral of the modular square of the timelike form factor
\beq && \vert \mathcal{F}_\pi(s) \vert^2 = \Theta(s_{\rm max} - s) \, \vert \mathcal{F}^{\rm data}_{\pi, {\rm Inter.}}(s) \vert^2 \nonumber \\
&& \hspace{1.3cm} + \, \Theta(s - s_{\rm max}) \,  \vert \mathcal{F}_\pi^{\rm pQCD}(s) \vert^2.  \label{eq:Fpi_DR2_modular} \eeq 
The data term is currently available from the threshold value $s_0 = 4m_\pi^2$ to $s_{\rm max} \sim 8.7$ GeV$^2$ \cite{BaBar:2012bdw}, 
meanwhile, the pQCD prediction is powerful in the large momentum transfers. 
Hence the modular dispersion relation is model independent. 

The state-of-art pQCD calculation has performed at NLO up to twist three and leading-order at twist four level \cite{Cheng:2020vwr}. 
As shown in Eq. (\ref{eq:ff-fact-kt}), the result includes contributions from different twist LCDAs, 
in which the high twist contributions are conceptually power suppressed by ${\cal O}(1/Q^2)$. 
For the form factors of pseudoscalar mesons, however, 
the terms proportional to two-particle twist three LCDAs are dominate in the intermediate and not too large momentum transfers ($Q^2 \sim {\cal O}(10,10^2)$) due to the chiral enhancement ${\cal O}( ( m_0^{\cal P} /\tilde{Q} )^2)$. 
Here the chiral mass $m_0^{\cal P} \sim {\cal O}(1)$ evolutes mildly on a hard scale, 
the effective longitudinal virtuality $\tilde{Q^2} \sim u_i Q^2$ in the hard scattering grows much slowly 
than the momentum transfers $Q^2$ itself due to the sudakov effect.

We now pick up the soft transversal dynamics which is overlooked in the previous pQCD predictions. 
For the goldstone pion, the terms proportional to quark mass can be safely neglected in the soft wave function Eq. (\ref{eq:iTMD}), 
the soft transversal dynamics is hence described solely by the iTMDs associated to the valence quark state. 
In order to obtain the transversal-size parameter $\beta_\pi$, 
we consider the constraint from the asymptotic behavior of double photon transition $\pi \to \gamma\gamma$ \cite{Li:2009pr,Kroll:2010bf}
\beq \beta^2_\pi = \frac{1}{8\pi^2f_\pi^2 \left(1 + a_2^\pi + a_4^\pi  + \cdots \right)}. \label{eq:TMD-t2-beta2} \eeq
Since the LQCD evaluations work well only for the lowest gegenbauer coefficient $a_2$ so far, 
we take the result obtained from the joint analysis of electromagnetic form factor 
with modular dispersion relation and precise LCSRs calculation \cite{Cheng:2020vwr}, 
say $a_2^\pi = 0.28 \pm 0.05, a_4^\pi = 0.19 \pm 0.06$. 
With the decay constant $f_\pi = 0.13$ GeV taken from Particle Data Group \cite{ParticleDataGroup:2024cfk}, 
we obtain $\beta^2_\pi = 0.51 \pm 0.04$ GeV$^{-2}$. 
This value corresponds to the mean transversal momentum $358 \pm 15$ MeV from Eq. (\ref{eq:iTMD-average}), 
revealing the soft transversal dynamics in the soft wave function. 

We do the fit of spacelike form factors obtained from the iTMDs-improved pQCD calculation to the modular dispersion relation, 
and obtain the chiral mass $m_0^\pi(1 \, {\rm GeV}) = 1.84 \pm 0.07 \, {\rm GeV}$. 
This value is thirty percents larger than the previous pQCD result \cite{Chai:2023htt} but consists with the ChPT \cite{Leutwyler:1996qg}, 
indicating a significient decrease of the form factor due to the soft transversal dynamics, especially in the small and intermediate momentum transfers. 
We notice that the pQCD prediction of the timelike form factor is embodied in the modular dispersion integral Eq. (\ref{eq:Fpi_DR2}) 
as the high energy tail contribution, in which the $m_0^\pi$ terms can not be separated out in the logarithm. 
So in the fit we firstly take an initial value of chiral mass ($1.6 \pm 0.4$ GeV) for the high energy tail contribution in the integrand, 
and do the numerical iteration to find the optional value of $m_0^\pi$ via the modular dispersion relation. 
The initial value of $m_0^\pi$ is the same as what we used in the previous pQCD study \cite{Chai:2023htt}, 
which we have determined by taking into account all the available results on the chiral mass of pion in the literatures. 

\begin{figure}[h] \begin{center} \vspace{-2mm}
\includegraphics[width=0.4\textwidth]{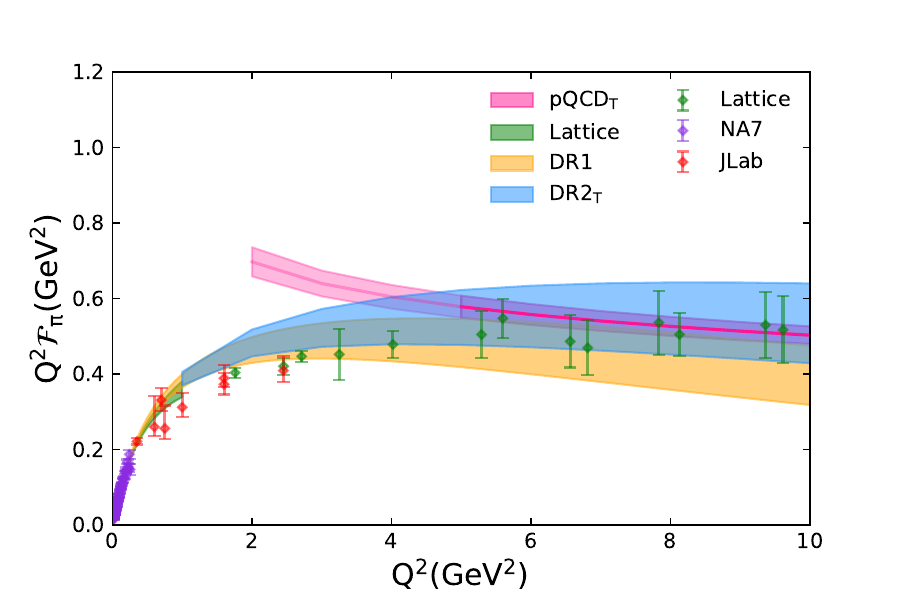} 
\end{center} \vspace{-6mm}
\caption{The iTMDs-improved pQCD prediction of pion electromagnetic form factor $Q^2F_\pi(Q^2)$. } 
\label{fig:pi-ff-itmd}
\end{figure}

In figure \ref{fig:pi-ff-itmd} we plot the iTMDs-improved pQCD prediction (magenta) of pion form factor $Q^2 F_\pi(Q^2)$, 
we compare it to the LQCD evaluation (green) \cite{Wang:2020nbf,Ding:2024lfj} and the measurements (purple and red) \cite{NA7:1986vav,JeffersonLabFpi-2:2006ysh,JeffersonLab:2008jve}, and find a well consistence between them. 
More impressively, we find that the power of pQCD prediction is improved down to a few GeV$^2$ after considering the iTMDs effect. 
In addition, we superpose the dispersion relation result in blue and yellow bands. 
The blue band is obtained by considering both the two terms in Eq. (\ref{eq:Fpi_DR2_modular}), 
the yellow band, however, is obtained by considering only the first term. 
We find a visible gap between them in the large momentum transfers. 
This is traced back to the logarithm expression in the modular dispersion relation Eq. (\ref{eq:Fpi_DR2}) 
that weights the high energy tail contribution. 

\begin{figure*}[t] \begin{center}  \vspace{-4mm}
\includegraphics[width=0.4\textwidth]{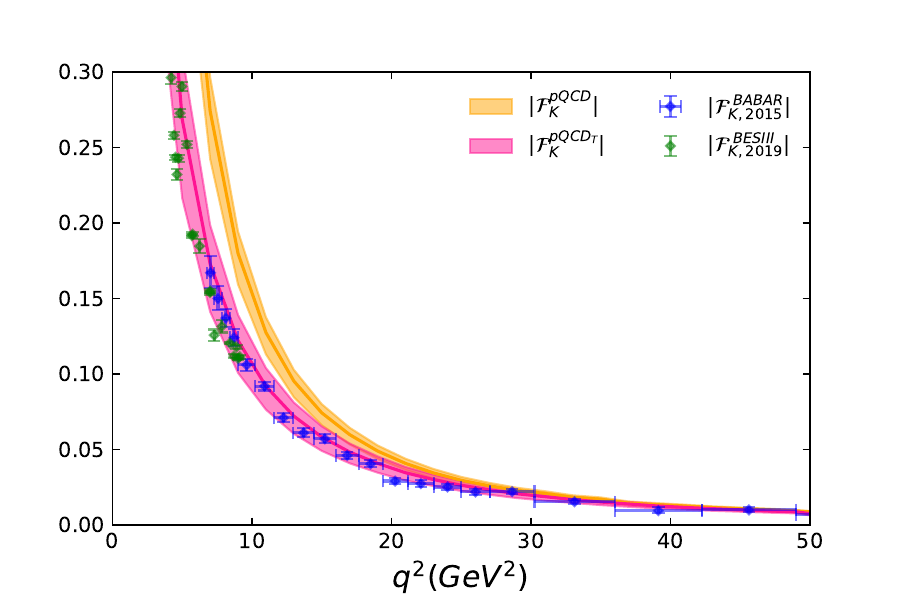} \hspace{8mm}
\includegraphics[width=0.4\textwidth]{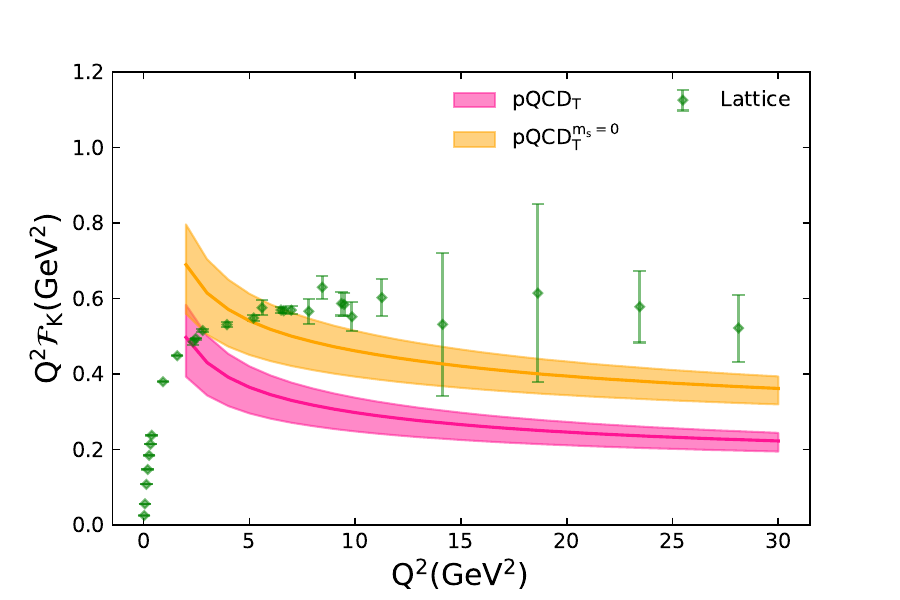} 
\end{center} \vspace{-6mm}
\caption{The pQCD predictions of kaon electromagnetic form factor in the timelike (left) and spacelike (right) regions. }
\label{fig:FF-kaon-pqcd-data}
\end{figure*} 

For the kaon electromagnetic form factor, there is no measurement in the spacelike region so far, 
but there are abundant data in the timelike region \cite{BaBar:2013jqz,BaBar:2015lgl,BESIII:2018ldc} 
spanning from the physical resonances to large momentum transfers. 
We can understand the soft transversal dynamics in kaon wave function by comparing the precise pQCD calculation directly to the timelike data. 

With the well-known ChPT relations \cite{Leutwyler:1996qg}, 
the chiral mass of kaon $m_0^K = 1.90^{+0.08}_{-0.12} \, {\rm GeV}$ is obtained without involving light quark masses. 
Here the uncertainty comes from the strange quark mass $\overline{m}_s(2 \, \mathrm{GeV}) = 96^{+8}_{-4} \, \mathrm{MeV}$. 
Other parameters in the kaon LCDAs are taken the same as in Refs. \cite{Ball:2006wn,Cheng:2019ruz}. 
The remaining unknown inputs are the transversal-size parameters $\beta_K^2$ and $\beta_K^{' 2}$ in the iTMDs. 
which could be obtained by fitting the pQCD prediction to the precise data. 
In the soft wave function of kaon, the term proportional to quark mass can not be ignored directly in Eq. (\ref{eq:iTMD}), 
but the three-particle contribution is still small ${\cal O}(m_s/m_0^K)$ in comparing to the valence quark term. 
In this case, the current accuracies of measurement and pQCD calculation prevent us to 
extract the transversal-size parameter $\beta_K^{' 2}$ associated to three-particle state. 
So we settle for the second best and take it equal to the transversal-size parameter associated to the valence quark state, say $\beta_K^{' 2} = \beta_K^2$. 
By fitting the iTMDs-improved pQCD prediction to the timelike data, we obtain $\beta_K^2 = 0.30 \pm 0.05$ GeV$^{-2}$. 
This value corresponds to the mean transversal momentum $0.55 \pm 0.07$ MeV for the valence quark state at leading twist. 

\begin{table}[t]\begin{center} \vspace{-2mm}
\caption{The mean transversal momenta and the conjugated distances of $\pi, K$ defined by Eq. (\ref{eq:iTMD-average}).}
\label{tab:transversal_meanvalue}
\begin{tabular}{c | c c c } \hline\hline
{\rm mean value} \quad & \quad $\phi$ \quad & \quad $\phi^p$  \quad & \quad $\phi^\sigma$ \quad  \nonumber\\ \hline
$\langle k_T^2 \rangle^{1/2}_\pi$ ({\rm GeV}) \quad & \quad $0.36 \pm 0.02$ \quad & \quad $0.40 \pm 0.02$ \quad & \quad $0.40 \pm 0.02$ \quad  \nonumber\\ 
$\langle b_T^2 \rangle^{1/2}_\pi$ ({\rm fm}) \quad & \quad $0.56 \pm 0.02$ \quad & \quad $0.50 \pm 0.02$ \quad & \quad $0.50 \pm 0.02$ \quad   \nonumber\\ \hline
$\langle k_T^2 \rangle^{1/2}_K$ ({\rm GeV}) \quad & \quad $0.55 \pm 0.07$ \quad & \quad $0.53 \pm 0.07$ \quad & \quad $0.52 \pm 0.07$ \quad   \nonumber\\ 
$\langle b_T^2 \rangle^{1/2}_K$ ({\rm fm}) \quad & \quad $0.37 \pm 0.05$ \quad & \quad $0.38 \pm 0.05$ \quad & \quad $0.39 \pm 0.05$ \quad   \nonumber\\  \hline\hline
\end{tabular}\end{center} \vspace{-2mm} \end{table}

We show the result of kaon electromagnetic form factor in figure \ref{fig:FF-kaon-pqcd-data}. 
For the timelike form factor, we compare the iTMDs-improved pQCD prediction (magenta) 
to the pQCD result without considering the soft transversal dynamics (yellow).  
We see that the iTMDs is indispensable to explain the data in the intermediate $q^2$. 
It also improves the power of pQCD prediction down to a few GeV$^2$ for the kaon form factor. 
We see from the right subgraph that the iTMDs-improved pQCD result (magenta) of spacelike form factor is small than the lattice data \cite{Ding:2024lfj}. 
In order to clarify the large $SU(3)$ flavor breaking which emerges an additional term proportional to $s$-quark mass $m_s$ 
in the twist three LCDAs, we depict the pQCD result with taking $m_s = 0$ in the yellow band.
We find the $m_s$-term decreases the kaon form factor by about thirty percents.
Our result shown in the magenta band agrees with the DSE approaches \cite{Yao:2024drm} and the collinear QCD factorization \cite{Chen:2024oem}. 

In table \ref{tab:transversal_meanvalue}, 
we list the mean transversal momenta and the conjugated distances associated to different LCDAs of $\pi$ and $K$. 
All the mean momenta on the transversal plane are found at soft scales.
The conjugated distances are smaller than their electric charge radius ($\langle r_\pi^2 \rangle^{1/2} = 0.67 \pm 0.08$ fm, 
$\langle r_K^2 \rangle^{1/2} = 0.56 \pm 0.03$ fm \cite{ParticleDataGroup:2024cfk}). 
The slight difference of the mean values between leading twist and twist three LCDAs comes from the three-particle contributions. 

\textbf{\textit{$k_T$ factorization revitalization.}}--With considering the soft transversal dynamics via iTMDs, 
the pQCD factorization formalism in Eq. (\ref{eq:ff-fact-kt}) is improved to 
\begin{widetext} \vspace{-3mm}
\beq &&{\cal F}(Q^2) = \sum_{t_1,t_2} \int du_i d b_i \, \psi^{(t_1)}(u_1, b_1, \mu_{r_1}) \otimes \psi^{(t_2)}(u_2, b_2, \mu_{r_2}) 
\otimes {\cal H}^{(t_1 t_2)}(u_i, b_i,\mu_f)  \times e^{-S(u_i, b_i, \mu_f)} . \label{eq:ff-fact-kt-iTMD} \eeq
\vspace{-5mm}\end{widetext}
Here $\psi(u_i,b_i,\mu)$ is the soft distribution amplitude of hadron formed outside the central scope of electromagnetic field, 
including both the longitudinal and transversal components as shown in Eq. (\ref{eq:pi-soft-DA}) and Eq. (\ref{eq:iTMD}). 

\begin{figure*}[t] \begin{center}  \vspace{-4mm}
\includegraphics[width=0.4\textwidth]{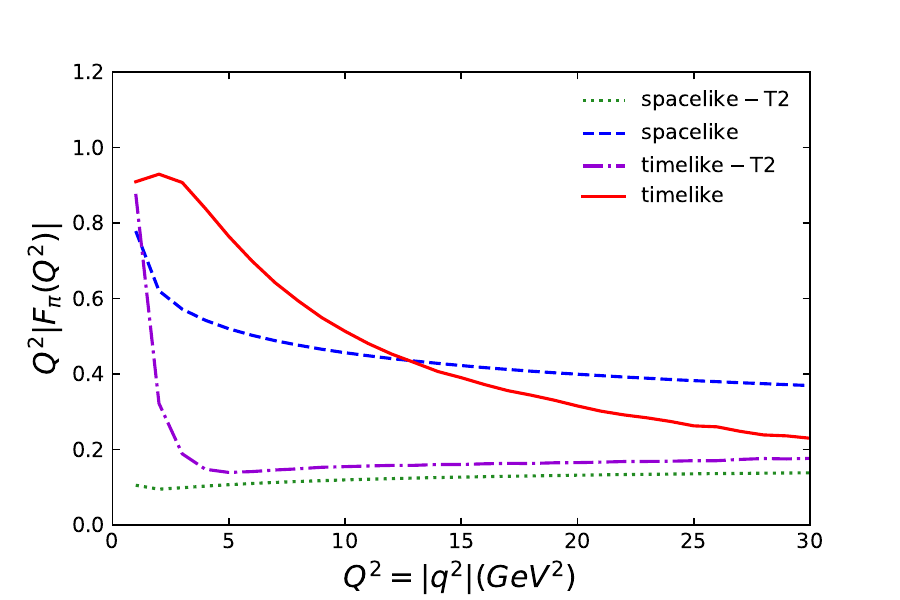} \hspace{8mm}
\includegraphics[width=0.4\textwidth]{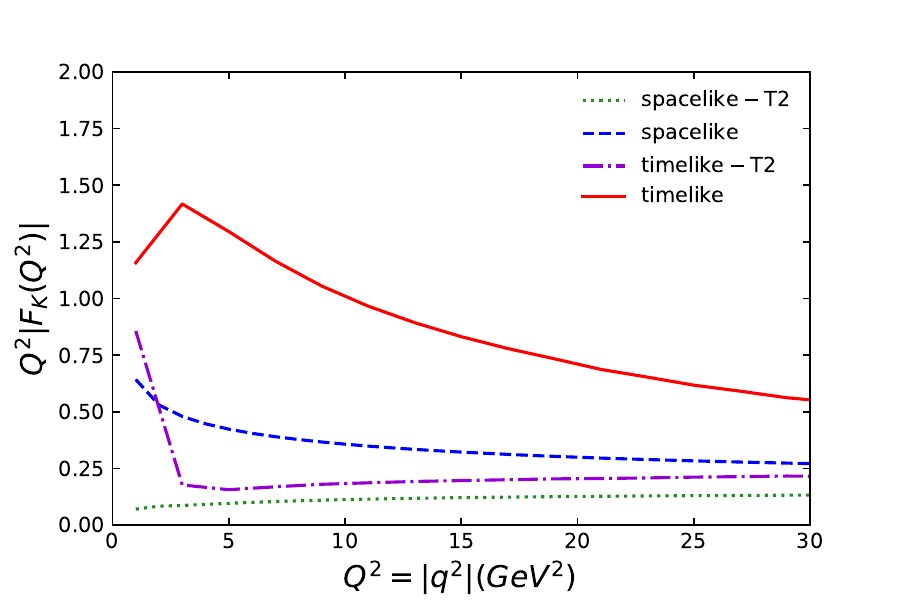}  
\end{center} \vspace{-6mm}
\caption{The state-of-the-art pQCD predictions of pion (left) and kaon (right) electromagnetic form factors.}
\label{fig:Q2F-sl-tl-pqcd}
\end{figure*} 
\begin{figure*}[t] \begin{center}  \vspace{-2mm}
\includegraphics[width=0.4\textwidth]{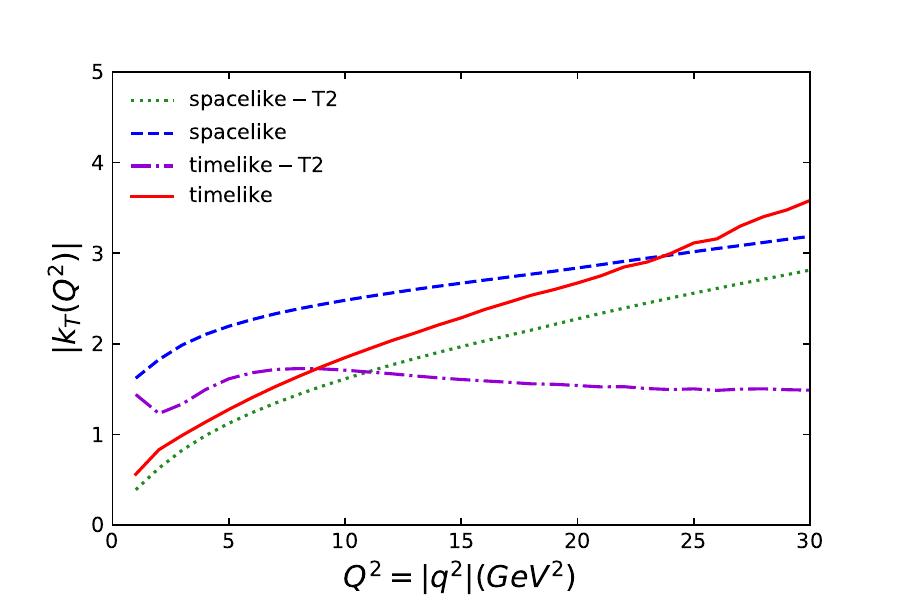} \hspace{8mm}
\includegraphics[width=0.4\textwidth]{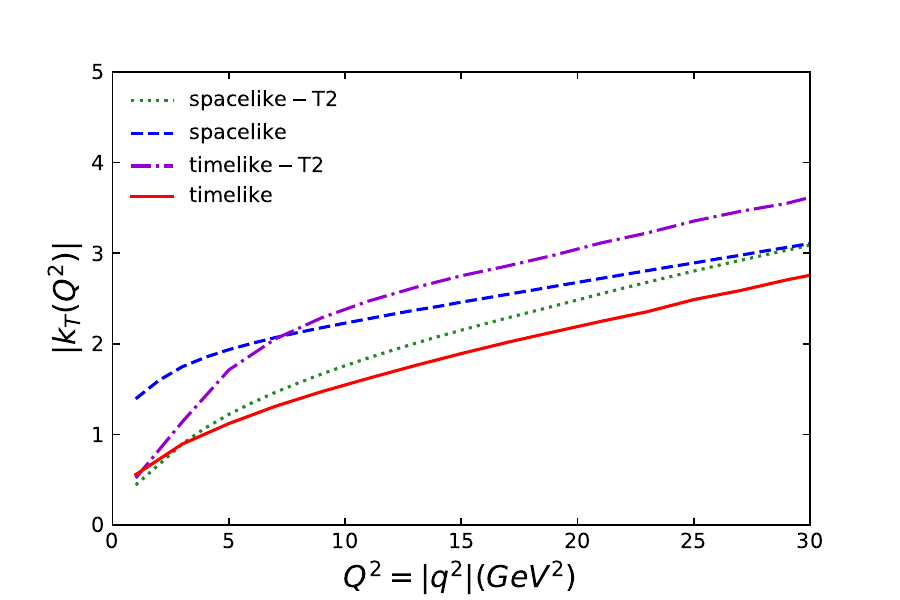}  
\end{center} \vspace{-6mm}
\caption{The mean transversal momenta in the pion (left) and kaon (right) electromagnetic form factors.}
\label{fig:kT-Fpi-sl-tl-pqcd}
\end{figure*} 

In figure \ref{fig:Q2F-sl-tl-pqcd} we show the state-of-the-art pQCD predictions of pion and kaon electromagnetic form factors. 
For the convenience of comparison, we depict both the spacelike and timelike results 
within a same horizontal coordinate $Q^2 = \vert q^2 \vert \in [4, 30]$ GeV$^2$. 
In order to show the high twist contributions (mainly from the chiral enhanced twist three LCDAs), 
we depict also the leading twist contributions (denoted "T2"). We find that 
(a) for the timelike pion form factor, the chiral enhancements decrease quickly with the growing of momentum transfers, 
and the leading twist contribution begins to dominate when $Q^2$ approaching to dozens GeV$^2$, 
(b) for the spacelike kaon form factor, the leading twist contribution begins to comparable to the high twists contribution at $Q^2 \sim 20$ GeV$^2$,
(c) for the spacelike pion form factors and the timelike kaon form factors, 
the chiral enhancements from twist three LCDAs hold to a much larger momentum transfers. 

These results can be understood from the factorization formalism in Eq. (\ref{eq:ff-fact-kt-iTMD}). 
The observable is a sum of contributions at different twists. They are written in
a convolution of the sudakov-multiplied hard amplitudes ${\cal H} e^{-S}$ with the soft hadron distribution functions $\psi$. 
Here ${\cal H} e^{-S}$ favors a small longitudinal fraction $u_i \sim {\cal O}(0.1)$ and a small transversal momentum (a few GeV), 
while $\psi$ highlights the moderate $u_i$ and large $k_T$. The result is finally a compromise between them. 
To clarify it quantitatively, we define the mean transversal momentum of electromagnetic form factor
\beq \vert k_{iT} \vert \equiv \left[ \sum_{t_1,t_2} \int du_i d k_{iT} \frac{k_{iT}^2 \psi^{(t_1)} {\cal H}^{(t_1 t_2)} 
e^{-S} \psi^{(t_2)}}{ \psi^{(t_1)} {\cal H}^{(t_1 t_2)} e^{-S}  \psi^{(t_2)}} \right]^{\frac{1}{2}} \label{eq:iTMD-average-ff} \eeq
and plot the result in figure \ref{fig:kT-Fpi-sl-tl-pqcd}. 
We see that \begin{itemize}
\item[(1)] The mean transversal momenta of spacelike form factors (depicted by the blue dashed curves) consistently reside at a hard scale, 
whereas the mean transverse momenta of timelike form factors (red thick curves) start to become hard 
at a large momentum transfer, specifically $Q^2 \sim 10$ GeV$^2$.
\item[(2)] In the electromagnetic interaction of pion, high twist terms lead to an increase in $\vert k_T \vert$ 
subsequently enhancing the longitudinal component of the propagator's virtuality within the hard scattering amplitude. 
Notably, the growth rate in the timelike interaction is much larger than that in the spacelike region, 
this explains the better convergence of twist expansion of timelike form factor, as evident in the left subgraph of figure \ref{fig:Q2F-sl-tl-pqcd}. 
\item[(3)] In the electromagnetic interaction of kaon, 
high twist contributions, conversely, reduce the mean transversal momentum of timelilke form factor by approximately $1$ GeV, 
leading to the the apparent chiral enhancement as illustrated in the right subgraph of figure \ref{fig:Q2F-sl-tl-pqcd}. 
Meanwhile, the high twist terms cause a slight increase in the mean transverse momentum of the spacelike form factor, 
which explains why the leading twist contribution is comparable to that of high twists in the spacelike kaon form factor. 
\end{itemize} 

From the perspective of physical imagery, the soft transversal oscillation inside pion (outside the scope of electromagnetic potential field) 
is highly excited by the electromagnetic interaction after the partons moving into the scope of potential field, especially in the timelike region. 
The interaction energy is subsequently transferred to the spectator quark via hard gluons,  
hence the Feynman picture of exclusive reaction converts completely to the hard mechanism. 
For the electromagnetic interaction of kaon, the energy transfers from the struck parton to the spectator parton too, 
the intensity of energy transfer, however, is lower than that in the pion due to the larger mass and inertia. 

\textbf{\textit{Summary.}}--In the exclusive QCD processes with large but not infinity momentum transfers, 
the hadrons are not moving as a rigid body because the partons inside is "oscillating" simultaneously in the transversal plane. 
In this letter, we propose an improved factorization for an exclusive QCD process, 
incorporating the soft transverse degree of freedom (described by iTMDs) missing in the $k_T$ factorization. 
The hard/hard-collinear dynamics are clearly embodied in the hard scattering amplitudes, 
meanwhile, the soft dynamics are absorbed into the iTMDs-associated LCDAs. 
With a combining analysis of electromagnetic form factors from the precise measurements, 
the iTMDs-improved pQCD predictions as well as the modular dispersion relation, 
we extract the transversal-size parameters in the gaussian expression of iTMDs. 
They are $\beta_\pi^2 = 0.51 \pm 0.04$ GeV$^{-2}$ and $\beta_K^2 = 0.30 \pm 0.05$ GeV$^{-2}$. 
We observe a significant decrease in electromagnetic form factors as a result of the iTMDs effect, 
especially when the momentum transfer is not excessively large. 
The updated result of chiral mass of pion, $m^\pi_0 = 1.84 \pm 0.07$ GeV, is consistent with the ChPT. 
The iTMDs-improved pQCD calculation also explains the precise measurement of timelike kaon form factor far away from the resonances very well. 
More impressively, it improves the pQCD prediction power of electromagnetic form factors down to a few GeV$^2$. 

We are grateful to Vladimir Braun, Heng-tong Ding, Jun Hua, Guang-shun Huang, Hsiang-nan Li, Wei Wang 
and Yu-ming Wang for fruitful discussions. 
This work is supported by the National Key R$\&$D Program of China under Contracts No. 2023YFA1606000 
and the National Science Foundation of China (NSFC) under Grant No. 11975112. 
J. C is also supported by the Launching Funding of Henan University of Technology (No.31401697).

\end{document}